**Vertical Transport of Subwavelength Localized Surface Electromagnetic Modes**

*Fei Gao[1], Zhen Gao[1], Youming Zhang[1], Xihang Shi[1], Zhaoju Yang[1], Baile Zhang[1,2],**

*Corresponding Author: E-mail: blzhang@ntu.edu.sg (B. Zhang)

[1]Division of Physics and Applied Physics, School of Physical and Mathematical Sciences, Nanyang Technological University, Singapore 637371, Singapore.
[2]Centre for Disruptive Photonic Technologies, Nanyang Technological University, Singapore 637371, Singapore.

## Abstract

Transport of subwavelength electromagnetic (EM) energy has been achieved through near-field coupling of highly confined surface EM modes supported by plasmonic nanoparticles, in a configuration usually staying on a two-dimensional (2D) substrate. Vertical transport of similar modes along the third dimension, on the other hand, can bring more flexibility in designs of functional photonic devices, but this phenomenon has not been observed in reality. In this paper, designer (or spoof) surface plasmon resonators ('plasmonic meta-atoms') are stacked in the direction vertical to their individual planes in demonstrating vertical transport of subwavelength localized surface EM modes. Dispersion relation of this vertical transport is determined from coupled mode theory and is verified with near-field transmission spectrum and field mapping with a microwave near-field scanning stage. This work extends the near-field coupled resonator optical waveguide (CROW) theory into the vertical direction, and may find applications in novel three-dimensional slow light structures, filters, and photonic circuits.

## 1. Introduction

Transport of subwavelength localized surface electromagnetic (EM) modes has been explored in various applications such as photonic energy transport [1-3], colourful super-resolution imaging [4], Dirac dispersions [5] and topological modes [6]. Such localized surface EM modes, also termed as localized surface plasmons (LSP) [7] at optical frequencies, are supported by metallic nanostructures [8-10] in a configuration usually on a two-dimensional (2D) substrate. Underlying these systems is the evanescent field coupling between neighbouring nanostructures. Many in-plane coupling of these localized EM modes can be accounted for with the theory of coupled resonator optical waveguide (CROW) [11-12], without considering the vertical coupling [11-15]. However, with the development of modern fabrication, multilayer photonic circuits become possible [16-17], making it desirable to study the transport of subwavelength localized surface EM modes through vertical coupling.

In this article, designer (or spoof) surface plasmon (DSP) resonators ('plasmonic meta-atoms') [18] are adopted to theoretically and experimentally demonstrate the vertical transport of subwavelength localized surface EM modes. Propagating DSPs are surface EM modes supported on extending subwavelength metal structures with surface plasmon-like dispersion relations in low frequencies ranging from microwave to infrared ranges [19-22], and localized DSPs can be supported on closed structured metal surfaces [23-24]. Subwavelength vertical transport of these localized modes is realized through stacking DSP resonators in the direction perpendicular to planes of individual DSP resonators. Starting from a single DSP resonator, three-dimensional (3D) field scanning shows that field profile along its vertical direction is in deep subwavelength scale, apart from its well-known in-plane subwavelength confinement that was confirmed previously [23-26]. Based on this evanescent field coupling, spectra measurement and field scanning are performed on a sandwiched free-standing sample of two DSP resonators that are stacked vertically. Subsequently, the coupling strength between a pair of coupled DSP resonators can be retrieved according to coupled mode theory [27]. Finally, for an infinitely long chain of these coupled DSP resonators in the vertical direction, theoretical calculation shows that the resonance frequency of a single DSP resonator can spread to a finite bandwidth and form a transport band. The dispersion of this new vertical transport band is verified via spectra measurement and field mapping on a sample of 5 vertically stacked DSP resonators.

## 2. Evanescent field along vertical direction

Field distributions along the vertical direction of an ultrathin DSP resonator has not been measured in the literature. In this work, firstly, the 3D field distribution is measured in the vicinity of a DSP resonator using a 3D field scanning stage. Fig. 1(a) shows the photo of the sample which consists of periodic grooves etched on a 18-μm-thick copper disk. The periodicity and groove width are $d = 1.256$ mm and $a = 0.628$ mm, respectively. The inner radius of the DSP resonator is $r = 3.000$ mm, and the outer radius is $R = 12.00$ mm. The ultrathin DSP resonator sticks to a 254.0-μm-thick Rogers RT5880 dielectric substrate which has relative permittivity 2.200+0.0009$i$. Since Teflon plates (relative permittivity 2.1) [28] will be used as spacers in later experiments in vertical coupling and mode transport, a Teflon plate of thickness $h = 6.0$ mm is placed under the RT5880 substrate in the field mapping. Two microwave monopoles, as the source and the probe, are employed at positions of the two dots close to the DSP resonator, as shown in Fig. 1(a). The measured near-field transmission spectrum is shown in Fig. 1(b), which shows evident cut-off frequency close to 6 *GHz* determined by $c/[4n_{eff}(R-r)]$ approximately [18], where *c* is light speed in free space, $n_{eff}$ is effective refraction index of environment and (*R-r*) is groove depth. The evident transmission peaks correspond to resonance modes of different orders, similar to previous works [23-26]. In the field mapping process, the probe scans the upper space above the DSP resonator to record $E_z$ distribution. Fig. 1(c) shows the 3D fields around the resonator for a hexapole mode (*f* = 5.32 GHz), the wavelength of which in free space is 56.39 *mm* much larger than groove period *d* = 1.256 *mm*. It confirms that the resonance modes supported by a DSP resonator are indeed subwavelength and highly confined in both the X-Y plane and along the Z direction. Other modes at different frequencies have similar profiles.

**3. Vertical coupling and mode splitting**

The vertical evanescent field provides vertical coupling. To demonstrate this vertical coupling, a second DSP resonator is attached onto the bottom of the Teflon plate on which the first DSP resonator locates, as shown in Fig. 2(a). Numerical simulation is first performed to study the near-field transmission spectrum. The excitation-probing configuration is similar to the experiment above. The two monopoles are located symmetrically in y = 0 plane with respect to resonator centres, but they are on different surfaces of the Teflon plate. The result of transmission in Fig.

2(b) shows that the original hexapole resonance peak (5.32 GHz) of a single DSP resonator splits into two new peaks at 5.10 GHz and 5.50 GHz because of the coupling of the two DSP resonators. (Another peak occurs near 5.60 GHz as one of the octopole modes.) This mode splitting is similar to the in-plane configuration [29], but here the coupling occurs in the vertical direction. Field distributions of $E_z$ component on planes 0.3 mm above the top and bottom surfaces of this two-resonator system at these two resonance frequencies are shown in Fig. 2(c-d). (Note that the lower field pattern that is 0.3 mm above the bottom surface is at a plane inside the Teflon layer.) In Fig. 2(c), the field pattern on the top surface has a $\pi$ phase shift compared with that on the bottom surface, while in Fig. 2(d) this phase shift vanishes. These different phase shifts can be further explained with field patterns at the y = 0 plane shown in Fig. 2(e-f). At 5.10 GHz, opposite charges are induced on top and bottom surfaces of the two-resonator system, which we refer to as out-of-phase mode (Fig. 2(e)). However, at 5.50 GHz, the same charge distribution are induced on both surfaces, which we refer to as in-phase mode (Fig. 2(f)).

Experimental measurement is then carried out to demonstrate the mode splitting induced by vertical coupling in the two-resonator system, as shown in Fig. 3(a). The measured spectrum in Fig. 3(b) matches well with the simulation results in Fig. 2(b) with slight frequency shifts of resonance peaks. The resonance peaks of the hexapole modes locate at 5.23 GHz and 5.49 GHz, respectively. The scanned field patterns at these resonance frequencies at 0.3-mm away from top and bottom surfaces of the two-resonator system are shown in Fig. 3(c-d). (Note that when scanning the field at the top surface, the probe is above the top surface. Yet when scanning the field at the bottom surface, the probe is below the bottom surface, and therefore the probed field will be opposite to the field above the bottom surface, the latter of which is used to compare with the field above the top surface. The reversing of the probe's orientation below the bottom surface can correct this sign change.) In Fig. 3(c), the field pattern in the left and right panels are left-right symmetric to each other (because of the reverse of x axis in the experiment) with a phase difference of π. It shows that at 5.23 GHz the phase shift of π for the field pattern is evident. At 5.49 GHz (Fig. 3(d)), no phase shift is observed for these field patterns. These observations match the simulation results in Fig. 2(e-f).

To quantitatively describe the vertical coupling, coupled mode theory [27] is employed to derive the coupling coefficient $\kappa$ as follows:

$$\begin{cases} -\frac{da_b}{dt} = i\omega_0 a_b + i\kappa\omega_0 a_t \\ -\frac{da_t}{dt} = i\omega_0 a_t + i\kappa\omega_0 a_b \end{cases} \quad (1)$$

where $a_b$ and $a_t$ denote fields in bottom and top resonators, respectively, and $\omega_0$ represents the hexapole resonance frequency of a single resonator (measured in Fig. 1(b)). Two normalized orthogonal eigen solutions of $\begin{bmatrix} a_b \\ a_t \end{bmatrix}$ in Eq. (1) are $\frac{1}{\sqrt{2}}\begin{bmatrix} 1 \\ -1 \end{bmatrix}$ and $\frac{1}{\sqrt{2}}\begin{bmatrix} 1 \\ 1 \end{bmatrix}$, corresponding to the out-of-phase mode in Fig. 3(c) and the in-phase mode in Fig. 3(d), respectively. Their corresponding eigen frequencies can be solved as $\omega_{-1} = \omega_0 - \kappa\omega_0$ (out of phase), and $\omega_1 = \omega_0 + \kappa\omega_0$ (in phase). After substituting $\omega_{-1}$ = 5.23 GHz, $\omega_1$ = 5.49 GHz, and $\omega_0$ = 5.32 GHz into above expressions, it can be easily obtained that $\kappa$ = 0.0244.

**4. Vertical transport of subwavelength localized DSP**

A one-dimensional (1D) infinite chain of stacked DSP resonators can be described by coupled mode theory as follows:

$$-\frac{da_n}{dt} = i\omega_0 a_n + i\kappa\omega_0 a_{n-1} + i\kappa\omega_0 a_{n+1} \tag{2}$$

where $a_n$, $a_{n-1}$ and $a_{n+1}$ denote fields in $n$-th, ($n$-1)-th and ($n$+1)-th resonators, respectively, as shown in the inset in Fig. 4(a). The periodicity along Z direction allows Bloch theorem $a_{n+1} = a_n e^{iKh}$ to apply, where $K$ represents the wavevector of vertical transport of a DSP mode, and $h$ is the inter-resonator distance. Then the intrinsic dispersion relation of the infinite chain can be obtained as $\omega = \omega_0\{1+2\kappa \cos(Kh)]$ as the curve in Fig. 4(a) which shows negative group velocity as [30, 31].

To experimentally demonstrate vertical transport of subwavelength DSP modes along a 1D chain, 5 vertically stacked DSP resonators are constructed with 6-mm-thick Teflon plates as spacers in-between, as shown in Fig. 4(b). Ideally, it would require an infinite number of resonators to form a continuous and smooth transmission band as in Fig. 4(a), which is the dispersion of an infinitely long 1D "bulk" system without terminals. In reality, only a finite chain with terminals can be adopted. Here we consider a chain of 5 resonators as sufficient to demonstrate this "bulk" property: the top and bottom resonators serve as terminals, and there are still multiple resonators (to be precise, 3 resonators) in-between. The inter-resonator distance 6 mm is much smaller than the working wavelength (about 56.39 mm) of hexapole mode. The measured near-field transmission spectrum, shown in Fig. 4(c), clearly shows a transmission band from 5.17 to 5.57 GHz around the original hexapole mode resonance frequency (Fig. 1(b)), being consistent with the bandwidth of dispersion curve in Fig. 4(a). As a comparison, the near-field transmission when the four lower resonators are removed is conducted, whose result in Fig. 4(c) shows that it is almost incapable of delivering EM energy from the bottom to the top of the structure. The scanned field pattern at 5.39 GHz on the top surface of the structure shows a clear hexapole mode profile. This confirms the transport of the hexapole mode from the bottom to the top through this vertical 5-resonator chain.

It should be noted that the phase of the field pattern at the top surface is out of phase compared to the field pattern at the bottom surface (not shown here), similar to the situation of two-resonator system (Fig. 3(c)). To show details of the wave propagation along this 5-resonator chain, numerical simulation is performed at 5.42 GHz for 5 planes, each of which is 0.3 mm above an individual resonator. As shown in Fig. 5, different resonators possess different phases along the propagation direction, and the top and bottom resonators are out of phase.

5. Summary

In conclusion, the vertical transport of subwavelength localized surface plasmon modes is demonstrated via stacking DSP resonators along the vertical direction. The evanescent field in the vertical direction around a DSP resonator causes mode splitting in a system of two vertically coupled DSP resonators, and forms a transport band for a chain of stacked DSP resonators. Based on the coupling strength retrieved with coupled mode theory, the dispersion relation of the vertical transport along a 1D infinite chain has been calculated. Transport capability and bandwidth have been experimentally demonstrated with a finite chain of 5 stacked DSP resonators. These results may find use for 3D slow light structures, filters, or photonic circuits.

**Acknowledgement**

This work was sponsored by Nanyang Technological University for Start-up Grants, Singapore Ministry of Education under Grant No. Tier 1 RG27/12 and Grant No. MOE2011-T3-1-005.

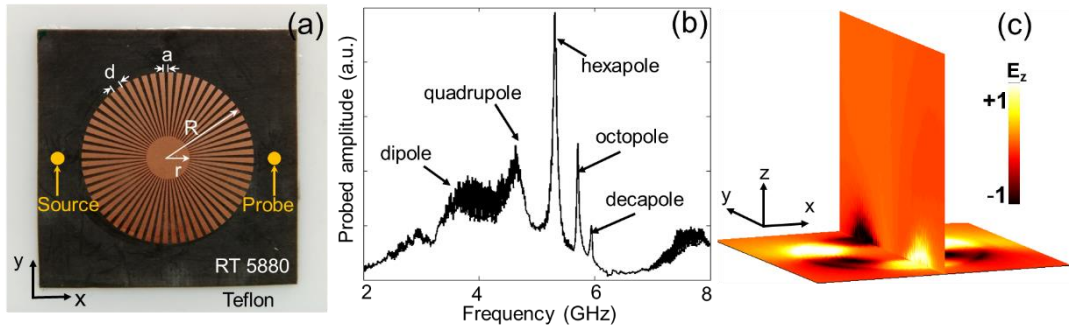

Figure 1. (a) Photo of a designer surface plasmon (DSP) resonator. Positions of near-field microwave source and probe are indicated by yellow dots. (b) Probed near-field transmission spectra. (c) Experimentally recorded $E_z$ field pattern on the X-Y plane and the Y-Z plane.

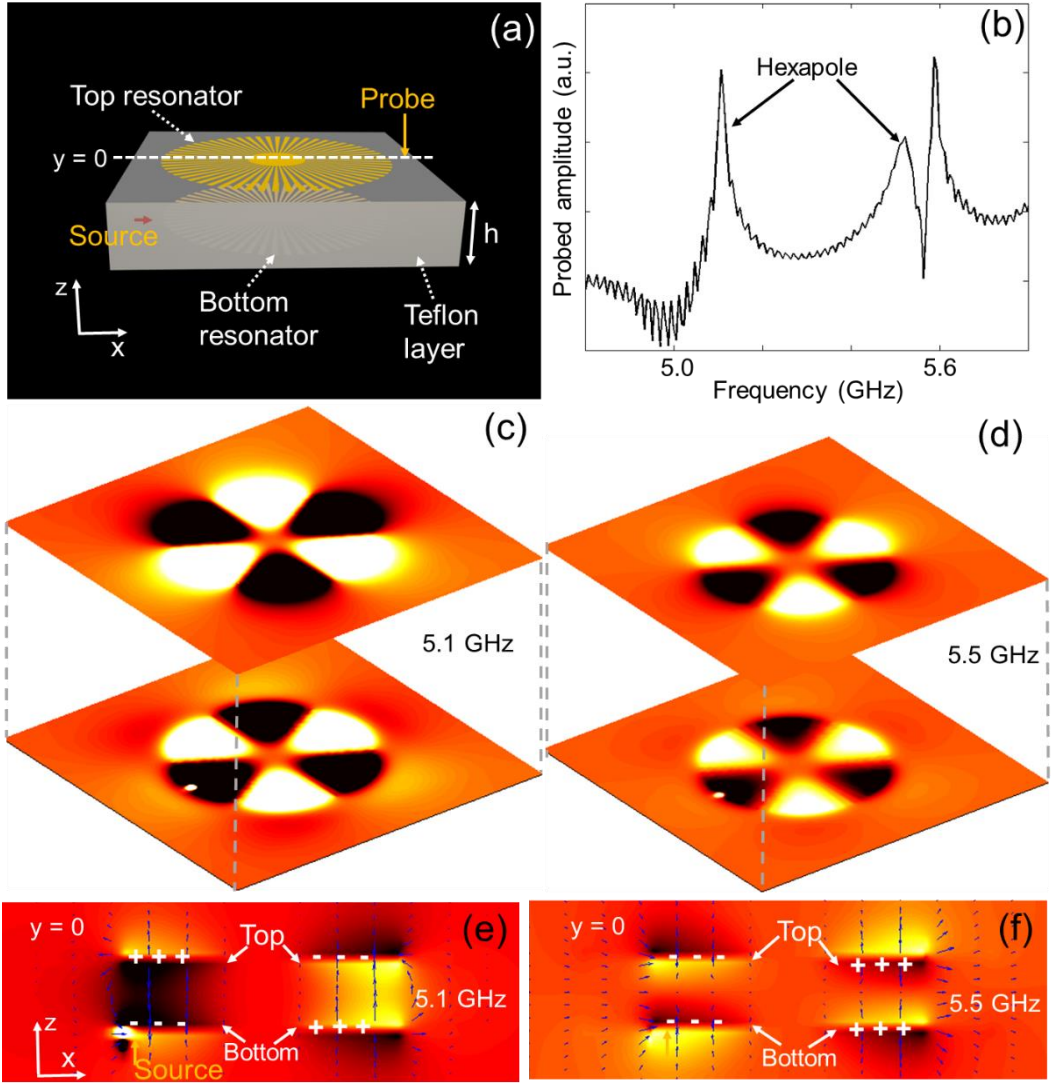

Figure 2. (a) Schematic of two vertically coupled DSP resonators. The source and the probe are on different surfaces. The white dashed line denotes y = 0. (b) Simulated near-field transmission spectrum. Two peaks indicated by black arrows are at 5.10 GHz and 5.50 GHz respectively. Both corresponds to a hexapole mode profile. (c)(d) $E_z$ field distribution in planes 0.3-mm above the top and bottom resonators at 5.10 GHz, and 5.50 GHz, respectively. (e) $E_z$ field distribution in the X-Z plane at 5.10 GHz. White '+' and '-' symbols denote accumulated positive and negative charges on top and bottom surfaces. Blue arrows represent $E$ field in the plane. (f) $E_z$ field distribution in the X-Z plane at 5.50 GHz.

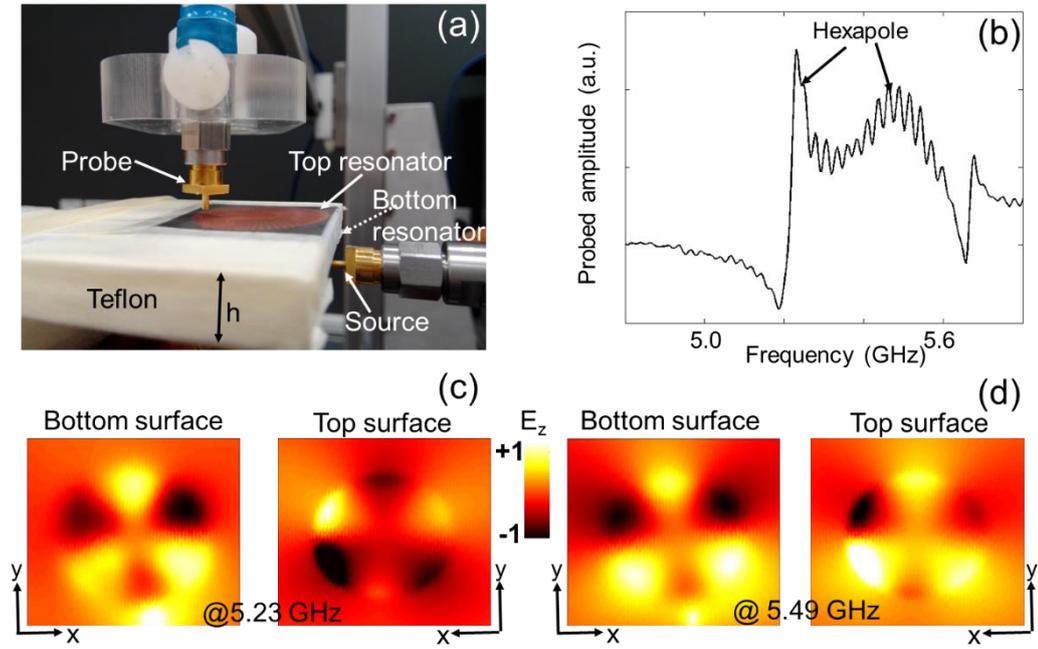

Figure 3. (a) Experimental set up. The teflon plate is long enough to support the two-resonator system. (b) Experimentally measured near-field transmission spectra. Two peaks indicated by black arrows are at 5.23 GHz and 5.49 GHz, respectively. (c) Detected $E_z$ field patterns at the top and bottom surfaces at 5.23 GHz show a phase difference of $\pi$. (d) No phase difference for these field patterns at 5.49 GHz.

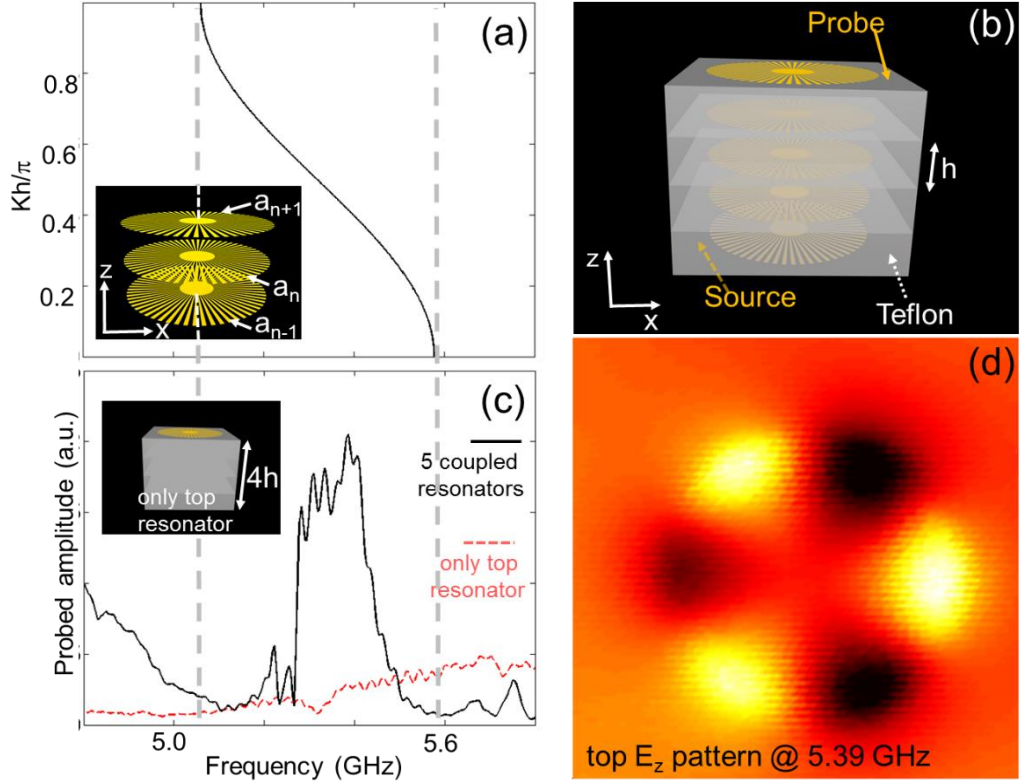

Figure 4. (a) Dispersion curve of the vertical transport of DSP modes along an infinite chain of DSP resonators. The inset shows three resonators of them. Teflon layers are not shown. (b) Schematic of 5 stacked DSP resonators separated by 4 Teflon layers with thickness $h$. (c) Detected near-field transmission spectra of the structure in (b). A control measurement on a sample as shown in inset (only top resonator and 4 Teflon layers) is also performed for comparison. (d) Detected $E_z$ field pattern on the top of the structure shown in (b).

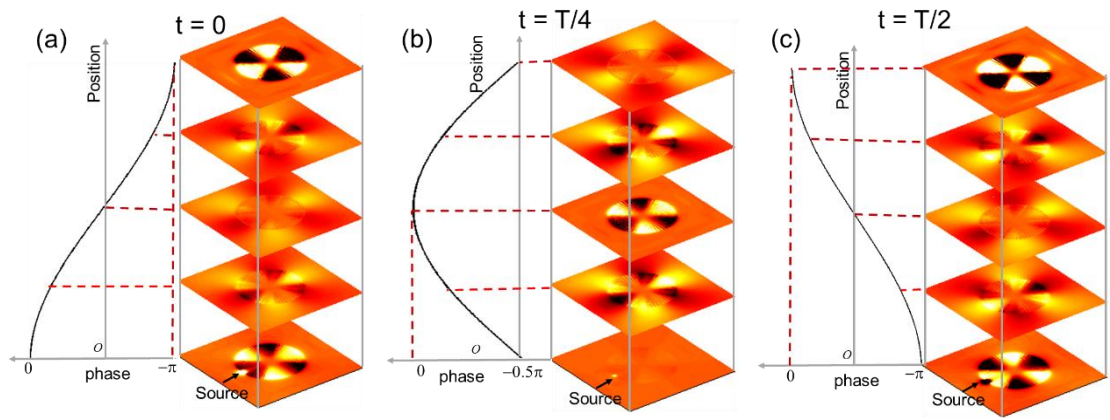

Figure 5. Field patterns of the 5-resonator chain at 5.42 GHz at time t = 0 (a), t = T/4 (b), and t = T/2 (c) respectively, where T is the period of field oscillation.